\documentclass[aps,prd,nofootinbib,reprint,superscriptaddress]{revtex4-1}
\usepackage{supertabular}
\usepackage{blindtext}
\usepackage{amsfonts}
\usepackage{tensor}
\usepackage{graphicx}
\usepackage{pict2e}
\usepackage{balance}

\usepackage[utf8]{inputenc}

\usepackage{amssymb, amsmath,environ}

\usepackage{color}


\newcommand{\bse}{\begin{subequations}}
	\newcommand{\ese}{\end{subequations}}
\newcommand{\be}{\begin{equation}}
	\newcommand{\ee}{\end{equation}}

\NewEnviron{eqsplit}{%
	\begin{equation}\begin{split}
			\BODY
	\end{split}\end{equation}
}

\makeatletter
\newcommand*\bigcdot{\mathpalette\bigcdot@{.5}}
\newcommand*\bigcdot@[2]{\mathbin{\vcenter{\hbox{\scalebox{#2}{$\m@th#1\bullet$}}}}}
\makeatother

\newcommand{\bea}{\begin{eqnarray}}
	\newcommand{\eea}{\end{eqnarray}}
\newcommand{\ba}{\begin{array}}
	\newcommand{\ea}{\end{array}}

\newcommand{\la}{\langle}
\newcommand{\ra}{\rangle}

\begin{document}
	
	\title{Impacts of isolated nucleon-nucleon correlations in relativistic $^{16}$O+$^{16}$O collisions}

	\begin{abstract}
		Nucleon-nucleon (NN) correlations in nuclear ground states can leave measurable imprints on the initial geometry of relativistic light-ion collisions. We introduce an Acceptance-Rejection Method (ARM) that provides a controlled framework for assessing the effects of selected two-body spatial correlations while keeping the underlying one-body radial density fixed. Starting from a common three-parameter Fermi (3pF) radial profile, we construct ensembles that separately incorporate pair-distance correlations ($\mathrm{ARM}_r$) and pairwise angular correlations ($\mathrm{ARM}_\theta$), and compare them with $^{16}\mathrm{O}$ configurations generated by Variational Monte Carlo (VMC) and Nuclear Lattice Effective Field Theory (NLEFT). We find that radial pair-distance correlations largely determine the bulk scale of initial-state geometric fluctuations, while angular correlations provide additional information required to reproduce eccentricity--energy covariance observables. The remaining small differences between the full many-body configurations and $\mathrm{ARM}_\theta$ indicate contributions from higher-order many-body correlations not encoded by two-body constraints alone. These results demonstrate how controlled correlation sampling can quantify the impact of nuclear many-body structure on initial-state observables in relativistic light-ion collisions.
	\end{abstract}
	
	\author{Qi Liu}
	\affiliation{School of Physics, Peking University, Beijing 100871, China}
	
	\author{Hadi Mehrabpour}
	\email{mehrabpour@fudan.edu.cn}
	\affiliation{School of Physics, Peking University, Beijing 100871, China}
	\affiliation{Center for High Energy Physics, Peking University, Beijing 100871, China}
	\affiliation{Institute of Modern Physics, Fudan University, Shanghai 200433, China}
	
	\author{Bing-Nan Lu}
	\email{bnlv@gscaep.ac.cn}
	\affiliation{Graduate School of China Academy of Engineering Physics, Beijing 100193, China}
	\maketitle

	\section{Introduction}\label{Introduction}
	
	Nucleons within the atomic nucleus do not behave as independent particles in a simple mean-field potential; rather, nuclear interactions generate nontrivial correlations in their spatial coordinates and momentum distributions~\cite{Freer:2017gip,Cruz-Torres:2019fum}. These NN correlations are an essential component of nuclear structure, encoding the interdependent spatial organization of protons and neutrons in the nuclear ground state~\cite{Cruz-Torres:2019fum,CLAS:2020mom}. In modern nuclear theory, \textit{ab-initio} frameworks construct light nuclei directly from realistic two- and three-nucleon forces~\cite{Piarulli:2019cqu,Piarulli:2022hml}, thereby generating correlated many-nucleon configurations that contain spatial structure beyond that described by conventional mean-field density profiles.
	
	The incorporation of \textit{ab-initio} nuclear configurations into the initial state of relativistic nuclear collisions~\cite{Giacalone:2025vxa} at facilities such as the Relativistic Heavy Ion Collider (RHIC) at Brookhaven National Laboratory~\cite{Huang:2023viw} and the Large Hadron Collider (LHC) at CERN~\cite{Citron:2018lsq,Brewer:2021kiv} provides a direct connection between low-energy nuclear structure and high-energy collective dynamics~\cite{Mehrabpour:2025rzt,Giacalone:2024luz,Zhang:2024vkh}. Recent studies have shown that spatial NN correlations and multi-nucleon clustering in light nuclei can leave measurable imprints on the initial collision geometry and, consequently, on final-state flow observables~\cite{Huang:2025uvc,Wang:2024vbb}. Understanding how different components of the nuclear spatial correlations propagate into collision observables is therefore important for establishing a quantitative connection between nuclear structure and relativistic heavy-ion phenomenology.
	
	In light nuclei such as $^{16}\text{O}$, multi-nucleon spatial correlations can manifest themselves through localized $\alpha$-cluster configurations~\cite{Broniowski:2013dia,Lin:2025dsm}. The geometric imprint of $\alpha$ clustering on initial-state spatial correlators in relativistic $^{16}\text{O}+^{16}\text{O}$ and $^{20}\text{Ne}+^{20}\text{Ne}$ collisions has been examined analytically in Ref.~\cite{hadi:2025}, while Ref.~\cite{Blaizot:2025scr} investigated the angular structure of multi-body correlations in $^{8}\text{Be}$. Furthermore, Ref.~\cite{Hu:2025eid} demonstrated that multi-nucleon cluster correlations can be isolated while keeping the one-body radial density fixed through Skyrme-DFT profiles with tetrahedral deformation. These studies motivate a more controlled investigation of how specific components of the correlated nuclear geometry influence observables in relativistic collisions.
	
	In this work, we introduce a reweighting framework based on an Acceptance-Rejection Method (ARM) that allows us to separately assess the effects of radial pair-distance and pairwise angular correlations in $N$-body nuclear configurations. Starting from a common, uncorrelated three-parameter Fermi (3pF) radial baseline, we construct controlled ensembles in which two-body spatial information is selectively imposed through radial and angular constraints. The resulting $\mathrm{ARM}_r$ and $\mathrm{ARM}_\theta$ ensembles provide a controlled way to quantify the contribution of different components of the nuclear spatial correlations to the initial-state collision geometry, without attempting to reconstruct the underlying many-body wave function in its entirety.
	
	We apply this framework to $^{16}\text{O}$ ground-state configurations generated by two distinct \textit{ab-initio} approaches: Nuclear Lattice Effective Field Theory (NLEFT)~\cite{Lu:2018bat,Elhatisari:2017eno} and Variational Monte Carlo (VMC)~\cite{Lonardoni:2017egu}. NLEFT employs a pionless effective field theory Hamiltonian on a spatial lattice and uses a pinhole algorithm to sample nucleon coordinates, whereas VMC solves the nuclear $A$-body problem using the Argonne v18 two-nucleon potential together with the Urbana X three-nucleon interaction. The central question addressed here is to what extent two-body radial and angular correlations are sufficient to reproduce the initial-state geometric fluctuations generated by the full many-body nuclear configurations. To address this question, we study two-particle eccentricities $\varepsilon_n\{2\}$, the total energy-per-entropy variance $\mathrm{var}(E/S)$, eccentricity--energy covariances $\mathrm{cov}(\varepsilon_n^2,\delta(E/S))$, and the four-particle cumulant ratio $\varepsilon_2\{4\}/\varepsilon_2\{2\}$. We find that radial pair-distance correlations largely determine the overall scale of the initial-state fluctuations, while angular correlations provide additional information needed to reproduce eccentricity--energy covariance observables. The remaining differences between the full many-body configurations and the two-body reconstructions provide sensitivity to higher-order many-body geometry not encoded by pairwise constraints alone.
	
	The remainder of this paper is organized as follows. In Sec.~\ref{Model setup}, we detail the ARM algorithm and define the initial-state observables. Section~\ref{results} presents the results for two-, three-, and four-particle correlators in $^{16}\text{O}+^{16}\text{O}$ collisions. Finally, Sec.~\ref{conclusion} summarizes our main findings and discusses their implications and outlook.

	\section{Methodology}\label{Model setup}
	
	An accurate description of single-particle and multi-particle nucleon distributions is essential for constructing the initial energy density profile in relativistic heavy-ion collisions~\cite{Alvioli:2009ab}. The ground-state single-particle (one-body) density profile $\rho(\mathbf{r})$ is formally obtained by integrating the $A$-nucleon ground-state wave function $\Psi(\mathbf{r}_1,\ldots,\mathbf{r}_A)$:
	\begin{equation}\label{q1}
		\rho(\mathbf{r}) =
		A \int \prod_{i=2}^{A} d\mathbf{r}_i\,
		|\Psi(\mathbf{r},\mathbf{r}_2,\ldots,\mathbf{r}_A)|^2.
	\end{equation}
	Beyond the one-body density, nuclear interactions generate correlations among the nucleons. Two-body spatial correlations can be characterized by the pair-distance distribution $g(\Delta r)$, which measures the probability density for finding two nucleons separated by the scalar distance
	\begin{equation}
		\Delta r = |\mathbf{r}_i-\mathbf{r}_j|.
	\end{equation}
	We quantify deviations from an uncorrelated reference through the pair-correlation measure~\cite{Alvioli:2009ab,Rybczynski:2019adt}
	\begin{equation}\label{eq:two}
		C(\Delta r) =
		1-\frac{g(\Delta r)}{g'(\Delta r)},
	\end{equation}
	where $g(\Delta r)$ denotes the pair-distance distribution of the correlated $A$-body configurations, while $g'(\Delta r)$ denotes the corresponding distribution obtained from an uncorrelated reference ensemble. With this convention, $C(\Delta r)>0$ corresponds to a local depletion of pair probability relative to the reference distribution, whereas $C(\Delta r)<0$ corresponds to an enhancement. The latter can arise from attractive interactions and/or spatial clustering, while the former can reflect short-distance repulsion and other mechanisms that suppress close nucleon pairs.
	
	To separately assess the effects of radial pair-distance correlations and pairwise angular correlations, we employ an Acceptance-Rejection Method (ARM)~\cite{Christensen}. We use a three-parameter Fermi (3pF) distribution as a common radial baseline,
	\begin{equation}\label{3pf}
		\rho_{\text{3pF}}(r) \propto
		\frac{1+w(r^2/R_0^2)}
		{1+\exp[(r-R_0)/a_0]},
	\end{equation}
	with $R_0=2.608\text{ fm}$, $a_0=0.513\text{ fm}$, and $w=-0.051$ for $^{16}\text{O}$~\cite{ANGELI201369}. Starting from this common one-body radial profile allows us to construct controlled reference ensembles in which selected two-body spatial correlations can be imposed while keeping the underlying radial baseline fixed. We consider two complementary implementations:
	
	\begin{enumerate}
		\item \textbf{Distance-Constrained ARM ($\text{ARM}_{r}$):}
		This procedure imposes the scalar pair-distance distribution while randomizing the angular orientation of the nucleons. Each nucleon's radial coordinate $r_i$ is sampled from $\rho_{\text{3pF}}(r)$, and candidate solid angles $\Omega_i=(\theta_i,\phi_i)$ are proposed uniformly. For a candidate position $\mathbf{r}_i$, the acceptance probability is
		\begin{equation}
			P_{\text{acc}} =
			\prod_{j<i}
			\frac{g(\Delta r_{ij})}
			{M_r\,g'(\Delta r_{ij})},
			\qquad
			M_r =
			\sup_{\Delta r}
			\left[
			\frac{g(\Delta r)}{g'(\Delta r)}
			\right],
		\end{equation}
		where $\Delta r_{ij}=|\mathbf{r}_i-\mathbf{r}_j|$. If the candidate is rejected, a new angular direction is sampled while keeping the radial coordinate fixed. In this way, $\text{ARM}_{r}$ reproduces the target radial pair-distance correlations while removing the specific angular correlations of the original configurations.
		
		\item \textbf{Angle-Constrained ARM ($\text{ARM}_{\theta}$):}
		This procedure imposes the pairwise angular distribution while maintaining the common 3pF radial baseline. The radial coordinates are again sampled from $\rho_{\text{3pF}}(r)$, while candidate angular directions are accepted according to the pair opening-angle distribution $g_{\text{ang}}(\cos\Theta_{ij})$ extracted from the corresponding \textit{ab-initio} configurations:
		\begin{equation}
			P_{\text{acc}} =
			\prod_{j<i}
			\frac{g_{\text{ang}}(\cos\Theta_{ij})}
			{M_\theta},
			\qquad
			M_\theta =
			\sup_{\cos\Theta}
			\left[
			g_{\text{ang}}(\cos\Theta)
			\right],
		\end{equation}
		where
		\begin{equation}
			\cos\Theta_{ij}
			=
			\hat{\mathbf r}_i\cdot\hat{\mathbf r}_j.
		\end{equation}
		Thus, $\text{ARM}_{\theta}$ retains the pairwise angular information of the original configurations while replacing their one-body radial structure by the common 3pF baseline. Any higher-order geometric structure that emerges in the resulting configurations is therefore a consequence of the imposed pairwise angular correlations rather than an explicitly imposed many-body constraint.
	\end{enumerate}
	
	\begin{figure}[t!]
		\begin{tabular}{c}
			\includegraphics[scale=.48]{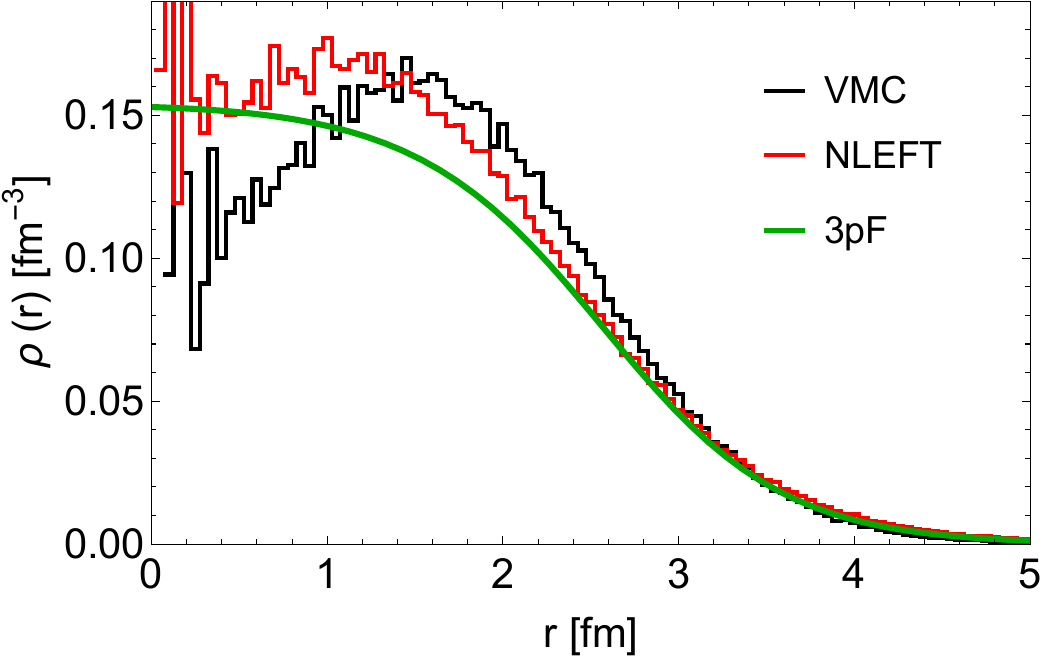}\\
			\includegraphics[scale=.48]{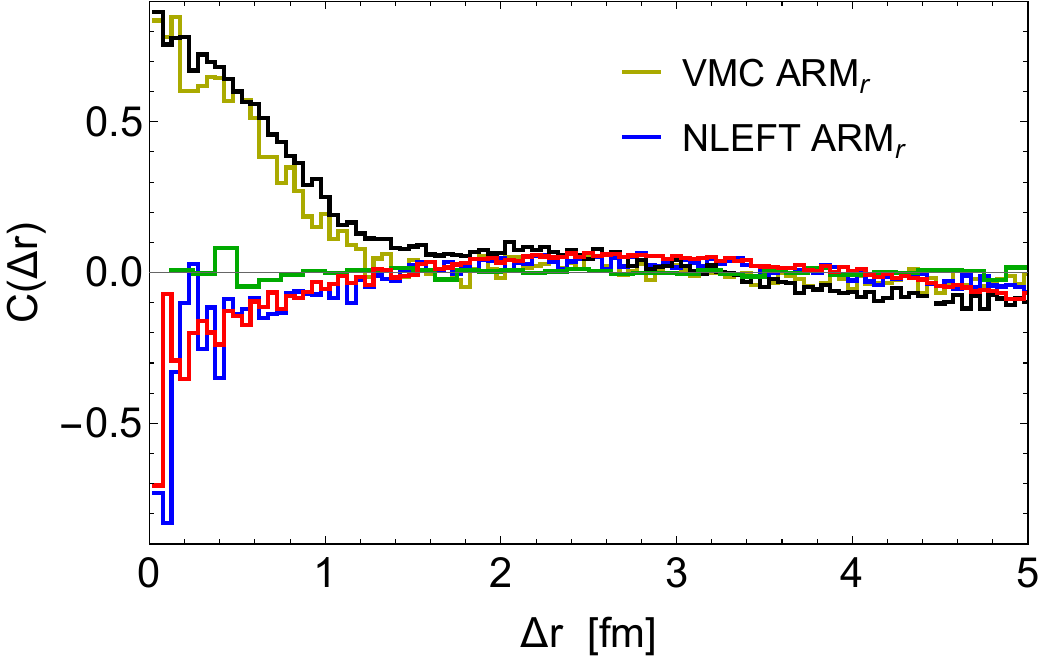}
		\end{tabular}
		\begin{picture}(0,0)
			\put(60,230){{\fontsize{15}{15}\selectfont \textcolor{black}{$^{16}$O}}}
			\put(-70,195){{\fontsize{12}{12}\selectfont \textcolor{black}{(a)}}}
			\put(-70,40){{\fontsize{12}{12}\selectfont \textcolor{black}{(b)}}}
		\end{picture}
		\caption{\label{fig1}
			Point-nucleon spatial distributions of $^{16}\text{O}$ across different initial-state configurations.
			(a) One-body radial density profiles $\rho(r)$ comparing VMC (black), NLEFT (red), and the 3pF baseline (green).
			(b) Two-body relative pair-correlation measures $C(\Delta r)$ comparing the original $N$-body quantum configurations (VMC: black, NLEFT: red), their radially reweighted counterparts ($\text{VMC ARM}_r$: yellow, $\text{NLEFT ARM}_r$: blue), and the uncorrelated 3pF baseline (green).}
	\end{figure}
	
	\textit{Radial Reweighting.--}
	To validate the radial reweighting scheme ($\text{ARM}_r$) and compare the spatial distributions of the input wave functions with the common radial baseline, we evaluate the point-nucleon spatial distributions of $^{16}\text{O}$ shown in Fig.~\ref{fig1}(a) presents the one-body radial density profile $\rho(r)$. The 3pF distribution [Eq.~\eqref{3pf}, green curve] provides a smooth reference profile, while the VMC (black) and NLEFT (red) configurations exhibit distinct radial structures. These differences reflect the different many-body ground-state configurations generated by the two \textit{ab-initio} approaches.
	
	Fig.~\ref{fig1}(b) presents the pair-correlation measure $C(\Delta r)$ defined in Eq.~\eqref{eq:two}. By construction, the uncorrelated 3pF reference corresponds to $C(\Delta r)=0$. The VMC configurations exhibit an enhancement of close nucleon pairs, resulting in a negative $C(\Delta r)$ at small separations. In contrast, NLEFT shows a depletion in the very short-distance region, corresponding to positive values of $C(\Delta r)$. These differences demonstrate that the two \textit{ab-initio} descriptions contain distinct short-distance pair structures.
	
	The $\text{ARM}_r$ configurations reproduce the corresponding pair-distance correlation measures of the VMC and NLEFT inputs within statistical uncertainties over the relevant range of $\Delta r$. At the same time, the radial reweighting procedure removes the specific angular organization of the original configurations. This makes $\text{ARM}_r$ a controlled reference with which to assess how much of the initial-state geometry can be attributed to radial pair-distance correlations alone.
	
	\begin{figure*}[t!]
		\begin{tabular}{c}
			\hspace*{-.3cm}\includegraphics[scale=.27]{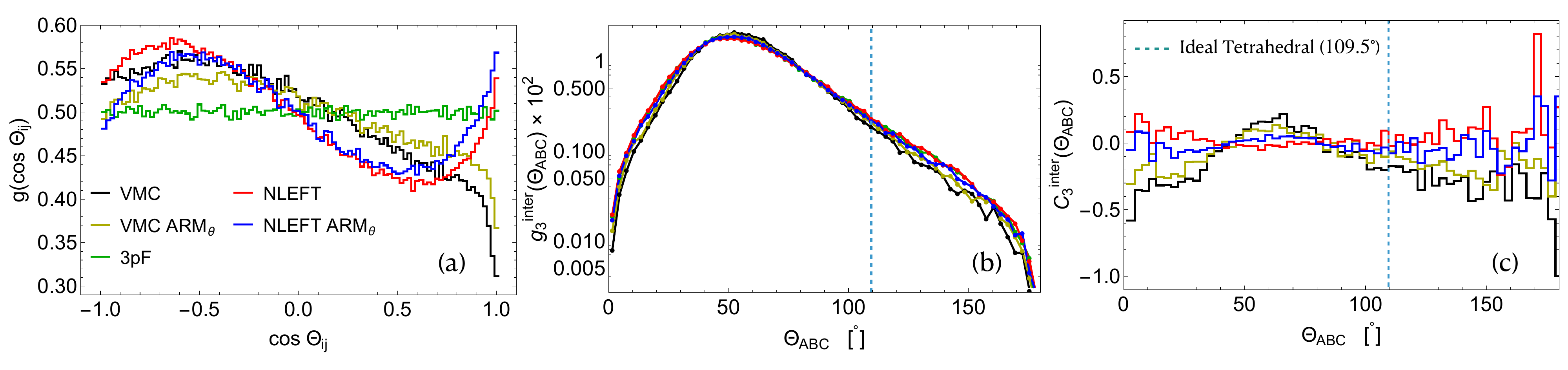}
		\end{tabular}
		\caption{\label{fig2}
			Relative pair-angular distributions and inter-cluster correlations for $^{16}\text{O}$.
			(a) Two-body pair opening-angle distribution $g(\cos\Theta_{ij})$ as a function of the relative polar-angle cosine $\cos\Theta_{ij}=\hat{\mathbf r}_i\cdot\hat{\mathbf r}_j$.
			(b) Raw three-cluster opening-angle distribution $g_3^{\text{inter}}(\Theta_{ABC})$ subtended at the center of mass of reconstructed $4\alpha$ clusters.
			(c) Three-cluster correlation function $C_3^{\text{inter}}(\Theta_{ABC})$, with the vertical dashed line denoting the ideal tetrahedral angle ($\Theta_{ABC}=109.5^\circ$).}
	\end{figure*}
	
	\textit{Angular Reweighting Validation.--}
	To validate the angle-constrained reweighting scheme
	($\mathrm{ARM}_{\theta}$), we examine the pair-angular and
	inter-cluster distributions shown in Fig.~\ref{fig2}.
	Panel~(a) presents the two-body pair opening-angle distribution
	$g(\cos\Theta_{ij})$, as defined in Eq.~(6). The uncorrelated
	3pF reference is isotropic, whereas the VMC and NLEFT
	configurations exhibit distinct angular structures. The
	angle-reweighted configurations, $\mathrm{VMC\ ARM}_{\theta}$
	and $\mathrm{NLEFT\ ARM}_{\theta}$, reproduce the corresponding
	features of their respective input angular distributions,
	demonstrating that the angular acceptance-rejection procedure
	preserves the targeted pairwise angular information.
	
	To investigate the corresponding higher-order cluster geometry,
	we partition the 16 nucleons of each $^{16}\mathrm{O}$ event
	into four candidate $\alpha$ clusters using $k$-means
	partitioning. The center of mass of cluster $A$ is denoted by
	$\mathbf{R}_{A}$, where
	$A\in\{1,2,3,4\}$. For any three distinct cluster centers,
	the inter-cluster opening angle is defined at one of the
	cluster centers. In particular, the angle at cluster $B$,
	formed by clusters $A$ and $C$, is
	\begin{equation}
		\cos\Theta_{ABC}
		=
		\frac{
			(\mathbf{R}_{A}-\mathbf{R}_{B})
			\cdot
			(\mathbf{R}_{C}-\mathbf{R}_{B})
		}{
			\left|\mathbf{R}_{A}-\mathbf{R}_{B}\right|
			\left|\mathbf{R}_{C}-\mathbf{R}_{B}\right|
		}.
	\end{equation}
	Here, $B$ denotes the vertex of the angle, while $A$ and $C$
	denote the two cluster centers connected to that vertex. The
	angle is evaluated only for three distinct cluster centers,
	with
	\[
	A\neq B,\qquad B\neq C,\qquad A\neq C.
	\]
	Thus, the four reconstructed cluster centers provide all
	ordered triples $(A,B,C)$ of distinct centers. The resulting
	angles are accumulated over the event ensemble.
	
	The normalized inter-cluster opening-angle distribution shown in
	Fig.~\ref{fig2}(b) is defined as
	\begin{equation}
		g_{3}^{\mathrm{inter}}(\Theta)
		=
		\frac{1}{N_{\mathrm{ang}}}
		\frac{dN_{\mathrm{ang}}}{d\Theta},
		\qquad
		\int_{0}^{\pi}
		g_{3}^{\mathrm{inter}}(\Theta)\,d\Theta=1,
	\end{equation}
	where $N_{\mathrm{ang}}$ denotes the total number of accepted
	inter-cluster angles in the event ensemble. The corresponding
	three-cluster correlation function shown in Fig.~\ref{fig2}(c)
	is defined relative to the uncorrelated 3pF reference as
	\begin{equation}
		C_{3}^{\mathrm{inter}}(\Theta)
		=
		\frac{
			g_{3}^{\mathrm{inter}}(\Theta)
		}{
			g_{3,\mathrm{3pF}}^{\mathrm{inter}}(\Theta)
		}-1,
	\end{equation}
	where $g_{3,\mathrm{3pF}}^{\mathrm{inter}}(\Theta)$ is obtained
	using the same cluster reconstruction, angle definition, and
	counting procedure applied to the uncorrelated 3pF configurations.
	Thus, $C_{3}^{\mathrm{inter}}=0$ corresponds to the 3pF reference,
	while positive and negative values indicate, respectively,
	enhancement and depletion of the inter-cluster angular
	distribution relative to that reference. For an ideal tetrahedral arrangement of four equivalent
	cluster centers, the angle between any two center-to-center
	directions meeting at a vertex is
	\begin{equation}
		\Theta_{\mathrm{tetrahedral}}
		=
		\arccos\!\left(-\frac{1}{3}\right)
		\simeq 109.5^{\circ},
	\end{equation}
	which is indicated by the vertical dashed line in
	Fig.~\ref{fig2}(b) and (c).
	
	Panels~(b) and (c) illustrate how the pairwise angular
	correlations manifest themselves in the higher-order geometry
	of the reconstructed four-cluster configuration. The original
	VMC configurations exhibit a distinct enhancement in the
	intermediate-angle region, while the corresponding
	$\mathrm{VMC\ ARM}_{\theta}$ configurations reproduce this
	characteristic feature. This demonstrates that nontrivial
	multi-cluster geometric structure can emerge from imposed
	pairwise angular correlations, even though no explicit
	three-body angular constraint is introduced. In contrast,
	NLEFT and $\mathrm{NLEFT\ ARM}_{\theta}$ exhibit a substantially
	weaker angular structure in the corresponding inter-cluster
	correlation measure.
	
	These results establish the role of the two ARM constructions
	as controlled reference ensembles:
	$\mathrm{ARM}_{r}$ isolates the contribution of radial
	pair-distance correlations, whereas $\mathrm{ARM}_{\theta}$
	tests how much of the nontrivial angular and higher-order
	geometry can emerge from pairwise angular information alone.
	
	To investigate how these spatial correlations affect relativistic collision observables, we simulate $^{16}\text{O}+^{16}\text{O}$ collisions at $\sqrt{s_{NN}}=200\text{ GeV}$~\cite{Huang:2023viw} using the TRENTo initial-condition model~\cite{Moreland:2014oya}. Centrality percentile classes are defined by sorting the simulated events according to their total integrated entropy,
	\begin{equation}
		S=\int s(x,y)\,dx\,dy,
	\end{equation}
	which is directly related to the final-state charged-particle multiplicity. In TRENTo, the mid-rapidity initial entropy density is parameterized through the generalized reduced thickness function
	\begin{equation}
		\frac{dS}{dy}\propto
		T_R(T_A,T_B)
		=
		\left(
		\frac{T_A^p+T_B^p}{2}
		\right)^{1/p},
	\end{equation}
	where $T_A$ and $T_B$ are the nuclear thickness functions of the two colliding nuclei. The parameter $p$ controls the nonlinear combination of the two thickness functions; $p=1$ corresponds to an arithmetic-mean prescription, while the limit $p\rightarrow0$ gives the geometric mean, $T_R=\sqrt{T_AT_B}$~\cite{Moreland:2014oya,Schenke:2012hg,Schenke:2016ksl}.
	
	We set $p=0$ throughout this work. This choice provides a smooth local mapping of the nuclear overlap geometry and minimizes additional nonlinear entropy-deposition effects, allowing the comparison to focus on differences originating from the nuclear configurations themselves ($3\text{pF}$, $\text{ARM}_r$, $\text{ARM}_\theta$, VMC, and NLEFT). We also performed calculations at LHC energy, $\sqrt{s_{NN}}=5.36\text{ TeV}$~\cite{Loizides:2025ule}, and found the same qualitative trends and geometric responses across the observables considered. We therefore present the $\sqrt{s_{NN}}=200\text{ GeV}$ results as representative of the high-energy regime.

	\begin{figure*}[t!]
		\begin{tabular}{c}
			\hspace*{-.3cm}\includegraphics[scale=.27]{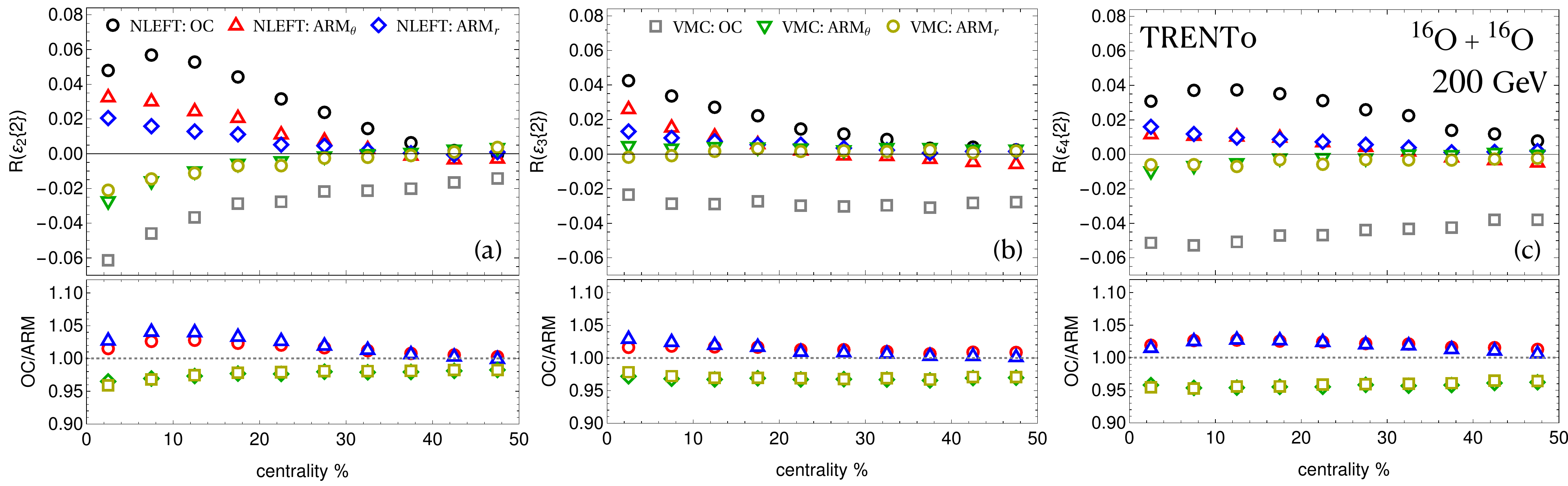}
		\end{tabular}
		\caption{
			Root-mean-square initial spatial eccentricities
			$\varepsilon_n\{2\}=\langle\varepsilon_n^2\rangle^{1/2}$
			for harmonic orders $n=2$ (panel a), $n=3$ (panel b), and $n=4$ (panel c) in $^{16}\text{O}+^{16}\text{O}$ collisions at $\sqrt{s_{NN}}=200\text{ GeV}$ as a function of centrality.
			The top subplots show the relative ratios
			$R(\varepsilon_n\{2\})=
			[\varepsilon_n\{2\}_{\mathrm{config}}/
			\varepsilon_n\{2\}_{\mathrm{3pF}}]-1$,
			comparing the original configurations (OC: NLEFT black circles, VMC grey circles) with their radially reweighted ($\mathrm{ARM}_r$: NLEFT blue diamonds, VMC yellow circles) and angle-reweighted ($\mathrm{ARM}_\theta$: NLEFT red triangles, VMC green triangles) counterparts.
			The bottom subplots show the direct reconstruction ratios OC/ARM, with
			$\mathrm{OC}/\mathrm{ARM}_\theta$ represented by NLEFT blue triangles and VMC yellow squares, and
			$\mathrm{OC}/\mathrm{ARM}_r$ by NLEFT red circles and VMC green squares.
		}
		\label{fig3}
	\end{figure*}
	\section{Correlators}\label{results}
	
	We now investigate how the spatial correlations encoded in the different nuclear configurations propagate into event-by-event fluctuations of the initial-state geometry and energy density. These initial-state quantities are connected to experimentally accessible final-state observables through the hydrodynamic response of the produced matter~\cite{Giacalone:2020dln,Gardim:2020sma,Samanta:2023amp}. At leading order, the relevant relations can be schematically written as
	\begin{align}
		\langle (\delta [p_T])^2 \rangle
		&\propto
		\mathrm{var}(E/\langle E\rangle),
		\label{q05}\\
		v_n\{2\}
		&\propto
		\varepsilon_n\{2\},
		\label{q06}\\
		\mathrm{cov}(v_n^2,\delta p_T)
		&\propto
		\mathrm{cov}(\varepsilon_n^2,\delta(E/S)),
		\label{q07}\\
		\frac{v_n\{4\}}{v_n\{2\}}
		&\approx
		\frac{\varepsilon_n\{4\}}{\varepsilon_n\{2\}},
		\label{q08}
	\end{align}
	where
	$
	\delta\mathcal{O}
	=
	\frac{\mathcal{O}-\langle\mathcal{O}\rangle}
	{\langle\mathcal{O}\rangle}.
	$
	These relations motivate the use of initial-state correlators as diagnostics of how well the ARM constructions retain the geometric information of the original \textit{ab-initio} configurations.
	
	Before examining the fluctuation observables, we first verify that the overall transverse size of the collision system is preserved by the reweighting procedure. We characterize the initial transverse size of the overlap region by~\cite{Bozek:2012fw,Bozek:2017elk}
	\begin{equation}
		R_s^2
		=
		2\frac{\int dx\,dy\,(x^2+y^2)s(x,y)}
		{\int dx\,dy\,s(x,y)},
		\label{eq:six}
	\end{equation}
	where $s(x,y)$ denotes the initial entropy density. The factor of two normalizes the root-mean-square radius such that $R_s$ coincides with the physical radius $R$ for a uniform entropy density distributed over a circular disk of radius $R$~\cite{Giacalone:2020dln}. In head-on collisions, $R_s$ is closely related to the mean transverse momentum and its event-by-event fluctuations~\cite{Giacalone:2020dln}. We find that the transverse size is preserved by the ARM constructions to within $1$--$2\%$ relative to the original configurations, confirming that the comparison of the fluctuation observables is not driven by a substantial change in the overall system size.

	\subsection{Two-Particle Eccentricities}
	
	The initial spatial anisotropies are characterized by the complex eccentricities
	\begin{equation}
		\mathcal{E}_n
		=
		\frac{
			\int dx\,dy\,(x+iy)^n s(x,y)
		}{
			\int dx\,dy\,(x^2+y^2)^{n/2}s(x,y)
		},
		\qquad
		\varepsilon_n=|\mathcal{E}_n|,
	\end{equation}
	which provide direct measures of the initial-state spatial geometry in relativistic heavy-ion collisions~\cite{Jia:2021qyu}. We quantify the magnitude of event-by-event eccentricity fluctuations through
	\begin{equation}
		\varepsilon_n\{2\}
		=
		\langle\varepsilon_n^2\rangle^{1/2},
	\end{equation}
	and compare each configuration with the uncorrelated 3pF baseline using
	\begin{equation}
		R(\varepsilon_n\{2\})
		=
		\frac{\varepsilon_n\{2\}_{\mathrm{config}}}
		{\varepsilon_n\{2\}_{\mathrm{3pF}}}
		-1.
	\end{equation}
	
	Figure~\ref{fig3} presents $R(\varepsilon_n\{2\})$ for $n=2$, $3$, and $4$ in $^{16}\text{O}+^{16}\text{O}$ collisions at $\sqrt{s_{NN}}=200\text{ GeV}$. The upper panels compare the original configurations with the $\mathrm{ARM}_r$ and $\mathrm{ARM}_\theta$ ensembles, while the lower panels show the corresponding reconstruction ratios.
	
	A central feature of the comparison is that all ARM ensembles are generated from the same 3pF radial baseline. Consequently, differences between the 3pF reference and the ARM ensembles arise from the imposed two-body spatial correlations rather than from differences in the underlying one-body radial density. Moreover, comparing $\mathrm{ARM}_r$ with $\mathrm{ARM}_\theta$ allows us to assess separately the contributions of pair-distance and pairwise angular correlations to the initial-state geometry.
	
	\begin{figure}[t!]
		\begin{tabular}{c}
			\hspace*{-.3cm}\includegraphics[scale=.27]{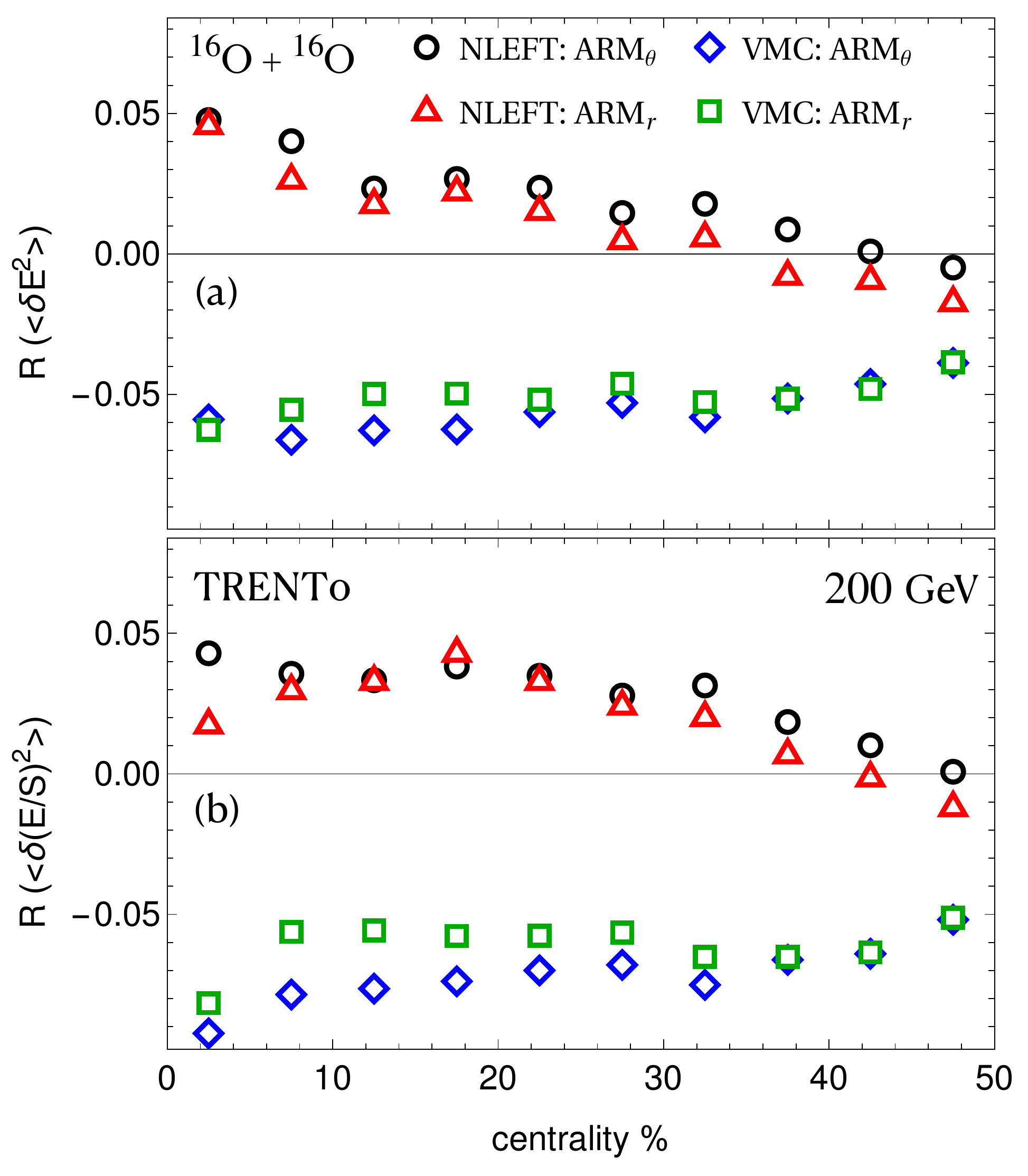}
		\end{tabular}
		\caption{
			Relative shifts in initial energy-density fluctuations in $^{16}\text{O}+^{16}\text{O}$ collisions at $\sqrt{s_{NN}}=200\text{ GeV}$ as a function of centrality.
			(a) Relative change in the total energy variance,
			$R(\langle\delta E^2\rangle)$.
			(b) Relative change in the energy-per-entropy variance,
			$R(\langle\delta(E/S)^2\rangle)$.
			Results are shown for NLEFT $\mathrm{ARM}_\theta$ (black circles), NLEFT $\mathrm{ARM}_r$ (red triangles), VMC $\mathrm{ARM}_\theta$ (blue diamonds), and VMC $\mathrm{ARM}_r$ (green squares) relative to the uncorrelated 3pF baseline.
		}
		\label{fig4}
	\end{figure}
	The original NLEFT and VMC configurations exhibit distinct responses relative to the uncorrelated 3pF reference. In NLEFT, the enhancement of close nucleon pairs seen in $C(\Delta r)$ is accompanied by enhanced spatial fluctuations and consequently larger values of $\varepsilon_n\{2\}$ in the most central collisions. In VMC, the suppression of very close pairs is accompanied by a reduction of the corresponding eccentricity fluctuations. These trends demonstrate that the short-distance pair structure contributes significantly to the magnitude of initial-state geometric fluctuations, although the eccentricities depend on the full spatial configuration and are not determined by $C(\Delta r)$ alone.
	
	The comparison between the two ARM constructions further clarifies the role of angular correlations. The radially constrained $\mathrm{ARM}_r$ ensembles already reproduce a substantial part of the shift away from the 3pF baseline. Thus, the scalar pair-distance distribution contains important information about the magnitude of the geometric fluctuations. Including the pairwise angular distribution in $\mathrm{ARM}_\theta$ moves the results further toward the original configurations, demonstrating that angular information provides an additional contribution to $\varepsilon_n\{2\}$.
	
	As shown in the lower panels of Fig.~\ref{fig3}, $\mathrm{ARM}_\theta$ reproduces the original configurations to within approximately $1$--$3\%$ for NLEFT and $2$--$4\%$ for VMC over the centrality range where the comparison is most sensitive to intrinsic nuclear structure. The remaining differences indicate that pairwise angular information does not contain the complete many-body geometry of the original configurations. In particular, three- and higher-body correlations can contain information that is not fixed by the corresponding two-body distributions. Toward more peripheral collisions, the macroscopic overlap geometry becomes increasingly important and the OC/ARM ratios approach unity.

	\subsection{Energy-Density Fluctuations}
	
	Beyond spatial eccentricities, nuclear spatial correlations also modify fluctuations of the total energy deposited in the collision zone. These fluctuations are related, through hydrodynamic response, to event-by-event fluctuations of the mean transverse momentum of final-state hadrons~\cite{Jia:2022ozr,Giacalone:2025vxa}. Defining the total initial energy as
	\begin{equation}
		E=\int dx\,dy\,\epsilon(x,y),
	\end{equation}
	the variance of the normalized energy can be written as
	\begin{equation}
		\la\delta E^2\ra\equiv\mathrm{var}(E/\langle E\rangle)
		=
		\frac{
			\langle E^2\rangle_{\mathrm{ev}}
			-
			\langle E\rangle_{\mathrm{ev}}^2
		}{
			\langle E\rangle_{\mathrm{ev}}^2
		},
	\end{equation}
	and is related to the mean-transverse-momentum fluctuations through Eq.~\eqref{q05}~\cite{Giacalone:2020dln,Blaizot:2014nia,Giacalone:2023hwk}.
	
	To investigate the effect of the nuclear correlations on the
	fluctuations of the deposited energy, Fig.~\ref{fig4} shows the
	relative deviation of the ARM results from the corresponding
	original configurations (OC),
	\begin{equation}
		R\!\left(\la\delta E^2\ra\right)
		=
		\frac{
			\la\delta E^2\ra_{\mathrm{ARM}}
		}{
			\la\delta E^2\ra_{\mathrm{OC}}
		}
		-1.
	\end{equation}
	Thus, $R=0$ corresponds to exact agreement with the original
	configuration, while positive (negative) values indicate that the
	ARM configuration produces larger (smaller) fluctuations than the
	corresponding OC result. Panel~(a) shows the relative deviation
	for the variance of the total deposited energy, while panel~(b)
	presents the corresponding result for $E/S$. The latter provides
	a useful measure in which fluctuations associated primarily with
	the overall collision volume are reduced.
	
		\begin{figure*}[t!]
		\begin{tabular}{c}
			\includegraphics[scale=.27]{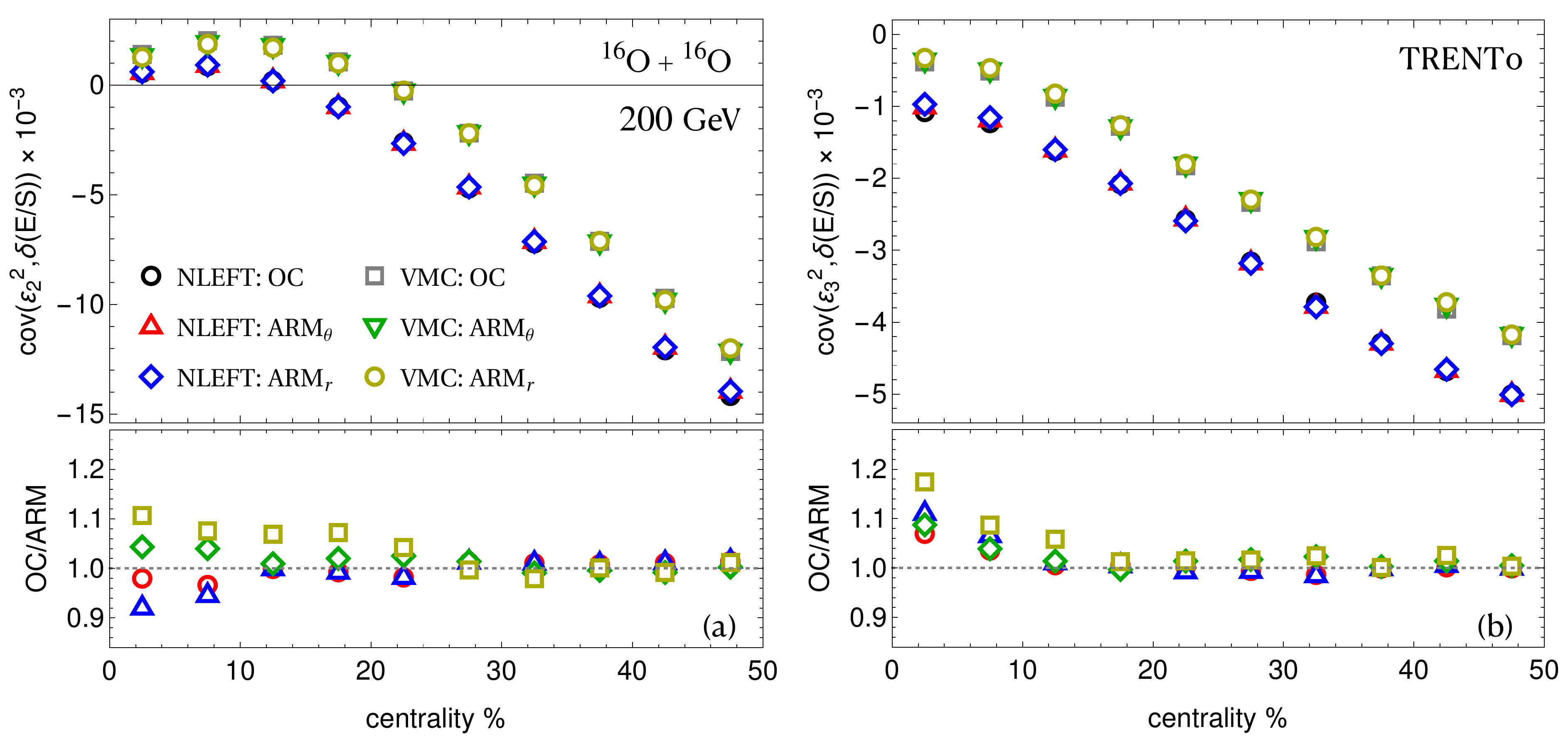}
		\end{tabular}
		\caption{
			Three-particle covariance between initial spatial eccentricity and energy-per-entropy fluctuations in $^{16}\text{O}+^{16}\text{O}$ collisions at $\sqrt{s_{NN}}=200\text{ GeV}$ as a function of centrality.
			(a) $\mathrm{cov}(\varepsilon_2^2,\delta(E/S))$.
			(b) $\mathrm{cov}(\varepsilon_3^2,\delta(E/S))$.
			Top subplots compare original configurations (OC: NLEFT black circles, VMC grey squares) with $\mathrm{ARM}_\theta$ (NLEFT red triangles, VMC green inverted triangles) and $\mathrm{ARM}_r$ (NLEFT blue diamonds, VMC yellow open circles).
			Bottom subplots show the corresponding reconstruction ratios.
		}
		\label{fig5}
	\end{figure*}
	The deviations of $\mathrm{ARM}_{r}$ from the OC results quantify
	the extent to which the radial pair-distance information alone
	reproduces the fluctuations of the original configurations.
	The comparison with $\mathrm{ARM}_{\theta}$ further shows how the
	inclusion of pairwise angular correlations modifies these
	fluctuations. The different behavior of the ARM constructions
	relative to OC demonstrates that the fluctuations of the
	deposited energy are sensitive not only to the radial
	nucleon--nucleon correlations but also to their angular
	organization. The persistence of the deviations in the
	$E/S$ observable indicates that the differences are not solely
	associated with fluctuations of the overall collision volume.

	\subsection{Three-Particle Correlators}
	
	Correlations between event-by-event anisotropic flow and mean transverse momentum provide sensitive probes of the connection between the initial energy-density profile and spatial eccentricities~\cite{Bozek:2016yoj,Bozek:2020drh,Giacalone:2023hwk}. Within the leading-order hydrodynamic response framework, the corresponding initial-state quantity is
	\begin{equation}
		\mathrm{cov}(v_n^2,\delta p_T)
		\propto
		\mathrm{cov}(\varepsilon_n^2,\delta(E/S)).
	\end{equation}
	We therefore study
	\begin{equation}
		\mathrm{cov}(\varepsilon_n^2,\delta(E/S))
		=
		\frac{
			\langle \varepsilon_n^2(E/S)\rangle_{\mathrm{ev}}
		}{
			\langle E/S\rangle_{\mathrm{ev}}
		}
		-
		\langle\varepsilon_n^2\rangle_{\mathrm{ev}},
	\end{equation}
	where the normalization by $\langle E/S\rangle_{\mathrm{ev}}$ follows from the definition of $\delta(E/S)$. This observable measures the event-by-event correlation between the magnitude of the spatial eccentricity and fluctuations in the energy deposited per unit entropy.
	
	Figure~\ref{fig5} presents
	$\mathrm{cov}(\varepsilon_2^2,\delta(E/S))$
	and
	$\mathrm{cov}(\varepsilon_3^2,\delta(E/S))$
	for $^{16}\text{O}+^{16}\text{O}$ collisions at $\sqrt{s_{NN}}=200\text{ GeV}$. The upper panels show the absolute covariance values, while the lower panels display the OC/ARM reconstruction ratios.
	
	For the elliptic covariance, both NLEFT and VMC give positive values in ultra-central collisions, where intrinsic nuclear geometry plays an important role. With increasing centrality, the elliptic geometry associated with the finite impact parameter becomes increasingly important and the covariance changes sign, becoming negative in mid-peripheral collisions. The difference between NLEFT and VMC in this region reflects the different spatial correlations present in their underlying nuclear configurations.
	
	The triangular covariance remains negative over the centrality range shown. The difference between the VMC and NLEFT results in the most central collisions reflects their different angular and cluster-like structures. In particular, the correlated angular organization of the VMC configurations leads to a smaller covariance magnitude than that obtained from NLEFT.
	
	The reconstruction ratios in the lower panels show that $\mathrm{ARM}_\theta$ reproduces the original covariances to within approximately $1$--$3\%$ for the centrality bins where the observable is well determined. The $\mathrm{ARM}_r$ construction captures the overall scale and centrality dependence but exhibits larger deviations, particularly in the most central collisions. This comparison demonstrates that radial pair-distance correlations determine an important part of the covariance, while pairwise angular correlations provide additional information required for a more accurate reconstruction.
	
	The reduction of the covariance magnitude when angular correlations are included can be understood in terms of the additional geometric constraints imposed on the nucleon configurations. In $\mathrm{ARM}_r$, the pair-distance distribution is reproduced while the angular arrangement retains greater freedom. Consequently, fluctuations of the eccentricity and the energy-density distribution can remain more strongly correlated. In $\mathrm{ARM}_\theta$, the pairwise angular structure further constrains the allowed event-by-event geometric rearrangements. This reduces the degree of independent variation between $\varepsilon_n^2$ and $E/S$ and consequently decreases the magnitude of their covariance.
	
	For $n=2$, this effect is particularly connected with the coherent quadrupole component of the angular structure, which accounts for part of the elliptic geometry. For $n=3$, the effect reflects the correlated multi-nucleon angular structure, including the cluster-like configurations present in the VMC ground state. Thus, the difference between $\mathrm{ARM}_r$ and $\mathrm{ARM}_\theta$ demonstrates that the pair-distance distribution alone does not contain all the information relevant for eccentricity--energy covariance observables.
	
	\begin{figure}[t!]
		\begin{tabular}{c}
			\includegraphics[scale=.27]{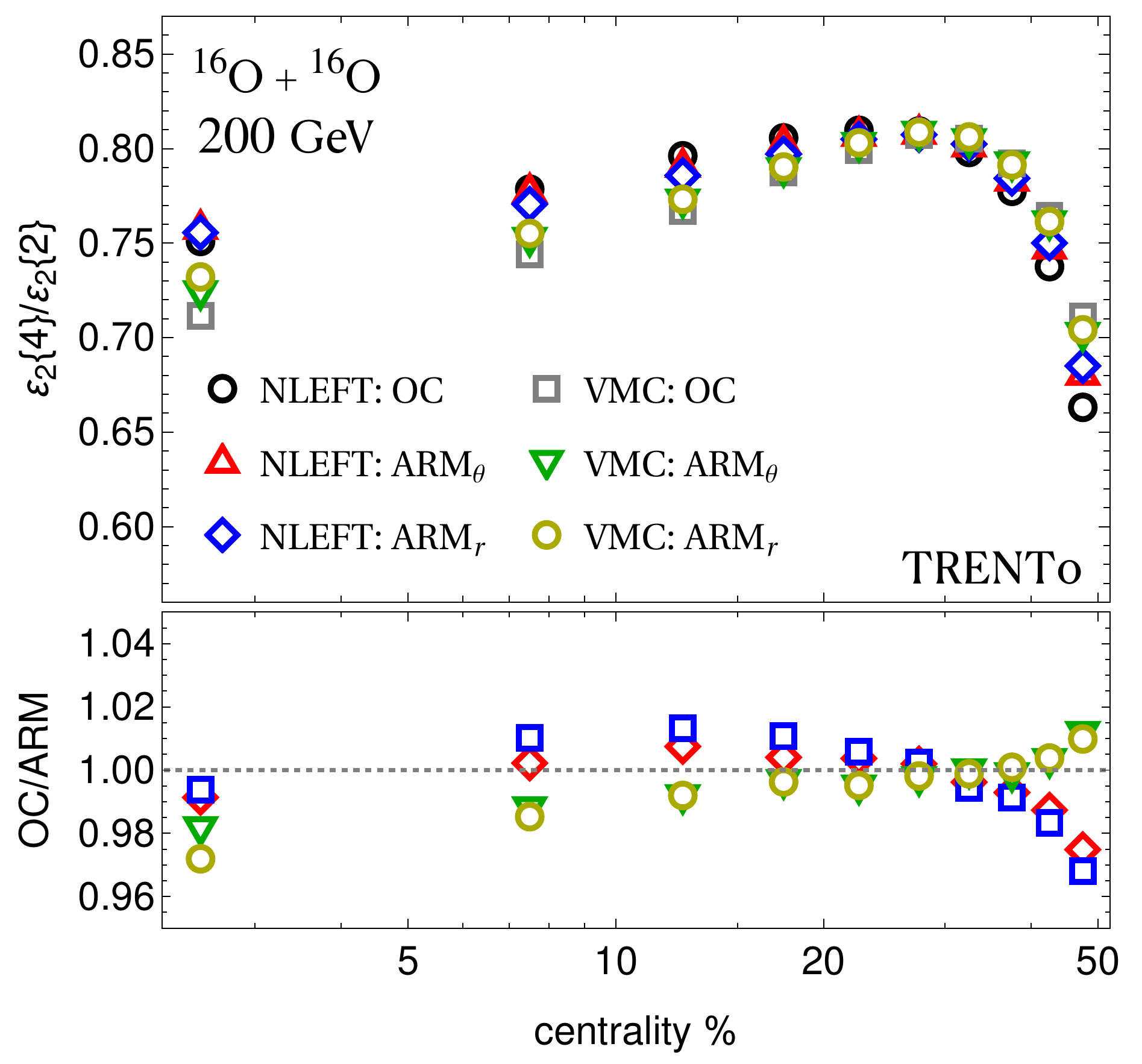}
		\end{tabular}
		\caption{
			Four-to-two-particle eccentricity cumulant ratio
			$\varepsilon_2\{4\}/\varepsilon_2\{2\}$ in $^{16}\text{O}+^{16}\text{O}$ collisions at $\sqrt{s_{NN}}=200\text{ GeV}$ as a function of centrality.
			The top panel compares the original configurations (OC: NLEFT black circles, VMC grey squares) with $\mathrm{ARM}_\theta$ (NLEFT red triangles, VMC green inverted triangles) and $\mathrm{ARM}_r$ (NLEFT blue diamonds, VMC yellow circles).
			The bottom panel shows the corresponding OC/ARM reconstruction ratios.
		}
		\label{fig6}
	\end{figure}
	
	\subsection{Four-Particle Eccentricity Cumulants}
	
	Although $^{16}\text{O}$ is doubly magic and near-spherical in mean-field descriptions~\cite{Delaroche:2009fa}, intrinsic many-body correlations can generate nontrivial geometric structures, including $\alpha$-cluster configurations~\cite{Giacalone:2024luz,Mehrabpour:2025rzt}. Their manifestation in higher-order eccentricity fluctuations provides a sensitive test of how much information about the full many-body nuclear geometry is retained by the two-body ARM constructions.
	
	We quantify non-Gaussian eccentricity fluctuations through the ratio
	\begin{equation}
		\varepsilon_2\{4\}/\varepsilon_2\{2\},
	\end{equation}
	where
	\begin{equation}
		\varepsilon_2\{2\}
		=
		\langle\varepsilon_2^2\rangle^{1/2},
		\qquad
		\varepsilon_2\{4\}
		=
		\left(
		2\langle\varepsilon_2^2\rangle^2
		-
		\langle\varepsilon_2^4\rangle
		\right)^{1/4}.
	\end{equation}
	Under approximate linear hydrodynamic response, the corresponding final-state flow ratio $v_2\{4\}/v_2\{2\}$ is related to the initial-state ratio $\varepsilon_2\{4\}/\varepsilon_2\{2\}$. Although nonlinear response and finite-number effects can become relevant in small systems, the initial-state cumulant ratio remains a useful measure of the non-Gaussian structure of the eccentricity distribution~\cite{Mehrabpour:2018kjs,Mehrabpour:2020wlu,Roubertie:2025qps,Alqahtani:2025vda,Mehrabpour:2023ign}.
	
	Figure~\ref{fig6} shows $\varepsilon_2\{4\}/\varepsilon_2\{2\}$ in $^{16}\text{O}+^{16}\text{O}$ collisions at $\sqrt{s_{NN}}=200\text{ GeV}$ up to $50\%$ centrality. The ratio exhibits a non-monotonic centrality dependence, increasing from approximately $0.70$--$0.75$ in ultra-central collisions to a broad maximum of about $0.81$ around $20$--$30\%$ centrality before decreasing in more peripheral collisions.
	
	The reconstruction ratios in the lower panel show that both $\mathrm{ARM}_r$ and $\mathrm{ARM}_\theta$ reproduce the original cumulant ratio to within approximately $1$--$3\%$ over most of the $0$--$40\%$ centrality range, with deviations remaining below approximately $4\%$. The strong performance of $\mathrm{ARM}_r$ indicates that the radial pair-distance information already constrains a substantial part of the width and non-Gaussian structure of the eccentricity distribution. The additional angular information in $\mathrm{ARM}_\theta$ provides further refinement, particularly in the regions where the original configurations contain stronger intrinsic angular structure.
	
	The success of the ARM constructions for this four-particle observable is significant because the cumulant depends on moments up to $\langle\varepsilon_2^4\rangle$. Nevertheless, the close agreement should not be interpreted as evidence that two-body correlations completely determine the full many-body nuclear wave function. Rather, it demonstrates that the selected two-body spatial correlations capture most of the information relevant to this particular higher-order geometric observable, while the residual differences quantify the contribution of geometric information not encoded in the imposed pairwise constraints.

	\section{Conclusion and Outlook}\label{conclusion}
	
	The intricacies of the nuclear force generate nontrivial many-body correlations of nucleons within nuclei that are essential to our understanding of nuclear structure and initial-state geometry in high-energy heavy-ion collisions. In this work, we investigated the impact of nucleon-nucleon spatial correlations encoded in ground-state $^{16}\text{O}$ configurations generated by two \textit{ab-initio} approaches, Nuclear Lattice Effective Field Theory and Variational Monte Carlo. Using an Acceptance-Rejection Method (ARM) based on a common three-parameter Fermi radial distribution, we separately reconstructed scalar pair-distance correlations and pairwise angular correlations, thereby isolating their respective contributions to the initial-state geometry of $^{16}\text{O}+^{16}\text{O}$ collisions.
	
	Our analysis of two-, three-, and four-particle geometric correlators leads to three main conclusions:
	
	\begin{enumerate}
		\item \textbf{Short-range $NN$ correlations control the direction of fluctuation changes.}
		The differences between the NLEFT and VMC pair-distance correlation functions at short separations are directly reflected in the initial-state fluctuations. The short-range pair enhancement in NLEFT, characterized by $C(\Delta r)<0$, produces localized spatial concentration and enhances the eccentricity fluctuations and energy-density variance relative to the uncorrelated 3pF baseline. Conversely, the short-range pair depletion in VMC, characterized by $C(\Delta r)>0$, suppresses local density fluctuations and reduces these observables. The same qualitative trend is observed in both $\varepsilon_n{2}$ and the energy-per-entropy variance, demonstrating that short-range pair correlations leave a measurable imprint on several classes of initial-state observables.
		
		\item \textbf{Pairwise angular correlations contain essential information beyond radial pair distances.}
		Because both ARM constructions are generated from the same 3pF radial baseline, their differences isolate the role of angular correlations in the nuclear configurations. The distance-constrained $\mathrm{ARM}_r$ reproduces the effects associated with the scalar pair-distance distribution, while the angle-constrained $\mathrm{ARM}_\theta$ incorporates the pairwise angular structure of the underlying \textit{ab-initio} configurations. The latter substantially improves the reconstruction of the original quantum configurations, reproducing the multi-particle covariance $\mathrm{cov}(\varepsilon_n^2,\delta(E/S))$ and the four-particle eccentricity ratio $\varepsilon_2\{4\}/\varepsilon_2\{2\}$ to within a few percent over the relevant centrality range. This demonstrates that the differences between the NLEFT and VMC initial states cannot be attributed to their one-body radial profiles alone.
		
		\item \textbf{Residual differences reveal higher-order many-body geometry.}
		Although $\mathrm{ARM}_\theta$ captures the dominant two-body angular information, it does not explicitly impose genuine three- and four-body correlations or the complete multipole structure of the nuclear wave function. The remaining $\sim2$--$4\%$ differences between the original quantum configurations and $\mathrm{ARM}_\theta$ therefore provide an estimate of the information that is not contained in pairwise distance and angular distributions alone. In particular, the comparison of the inter-cluster geometry indicates that higher-order correlations associated with the $4\alpha$ structure of $^{16}\text{O}$ can contribute beyond the information encoded in two-body angular correlations.
		
	\end{enumerate}
	
	The analysis also demonstrates the importance of using appropriate statistical and geometric treatments when studying ultra-central light-ion collisions. Following the referee's suggestion, we increased the initial-state statistics through random three-dimensional spatial rotations of the original configurations and the ARM samples, thereby generating large event ensembles without altering their intrinsic correlations. In addition, using the volume-corrected quantity $\delta(E/S)$ rather than $\delta E$ reduces trivial volume fluctuations and provides a more stable measure of the correlation between eccentricity and energy-density fluctuations in ultra-central events. These improvements remove the previously observed statistical irregularities and yield a smooth centrality dependence of the three-particle covariances.
	
	Overall, our results show that the Acceptance-Rejection framework provides a controlled way to decompose the geometric information carried by nuclear many-body configurations into radial pair-distance and angular components. Short-range pair correlations primarily determine the direction of fluctuation changes, while angular correlations are essential for recovering the detailed geometry of the \textit{ab-initio} configurations. At the same time, the residual differences between the full quantum configurations and pairwise-reweighted samples demonstrate the limitations of a purely two-body description and point toward genuine higher-order many-body structure.
	
	The present study focuses on initial-state geometry and therefore does not attempt to establish a direct quantitative mapping between $\varepsilon_2\{4\}/\varepsilon_2\{2\}$ and the experimentally measured $v_2\{4\}/v_2\{2\}$. In small collision systems such as $^{16}\text{O}+^{16}\text{O}$, nonlinear hydrodynamic response and finite-number effects can modify this mapping. A quantitative comparison with final-state observables will require dedicated viscous hydrodynamic calculations including these response effects.
	
	As a natural extension, future work will incorporate explicit three- and four-body density constraints into the sampling framework. Applying the method to other light and clustered nuclei, such as $^{12}\text{C}$, and $^{20}\text{Ne}$, will provide a systematic way to investigate how higher-order nuclear structure is transmitted to initial-state geometry and, ultimately, to observables in relativistic light-ion collisions.
	
		\begin{figure*}[t!]
		\begin{tabular}{c}
			\includegraphics[scale=.27]{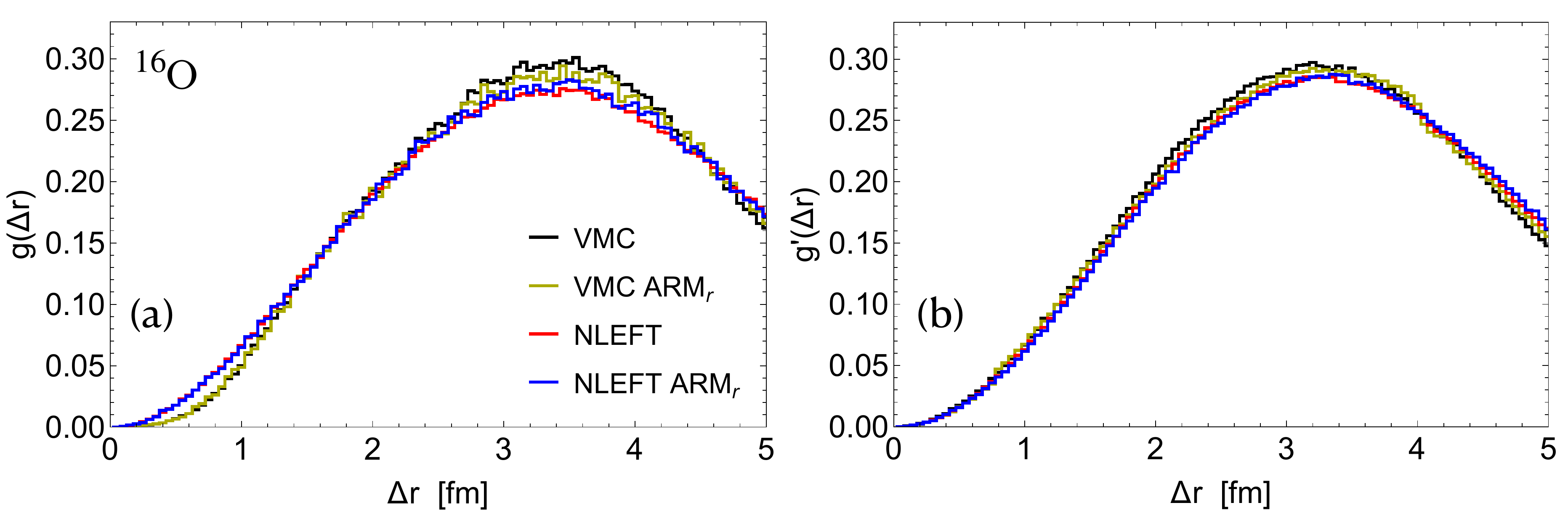}
		\end{tabular}
		\caption{Nucleon-nucleon pair-distance probability distributions $g(\Delta r)$ in $^{16}\text{O}$ ground states calculated from VMC and NLEFT. (a) Unnormalized pair-distance distributions $g(\Delta r)$ comparing the original quantum configurations (VMC: black line, NLEFT: red line) with the corresponding radially reweighted Acceptance-Rejection Method results ($\text{VMC ARM}_r$: olive line, $\text{NLEFT ARM}_r$: blue line). (b) Normalized pair-distance distributions $g'(\Delta r)$, illustrating the common overall radial shape obtained after normalization for the two \textit{ab-initio} frameworks.} 
		\label{fig_A1}
	\end{figure*}
	
	\section*{Acknowledgements}
	We thank Jean-Paul Blaizot and participants of the YITP workshop for fruitful discussions and comments on improving the physics of this paper. H.M is supported in part by the National Natural Science Foundation of China under Grant No. 12247107. B.N.Lu is also supported by the National Security Academic Fund (Grant No.U2330401).
	
	\appendix
	\section{Pair-Distance Probability Distributions in $^{16}\text{O}$}
	\label{app:pair_distributions}
	
	To evaluate the fidelity of the Acceptance-Rejection Method ($\text{ARM}_r$) in reproducing the two-body spatial geometry, we examine the pair-distance probability density distribution $g(\Delta r)$ for $^{16}\text{O}$ configurations generated by VMC and NLEFT. 
	In Fig.~\ref{fig_A1}(a), we present the unnormalized pair-distance distributions $g(\Delta r)$ for VMC (black curve) and NLEFT (red curve), together with their corresponding $\text{ARM}_r$ reconstructions ($\text{VMC ARM}_r$: olive curve; $\text{NLEFT ARM}_r$: blue curve). At short distances ($\Delta r<1\,\mathrm{fm}$), the two \textit{ab-initio} configurations exhibit distinct pair-distance distributions. NLEFT shows an enhanced probability density at small $\Delta r$, while VMC exhibits a suppressed probability as $\Delta r\rightarrow0$, reflecting the different short-distance correlation structures generated by the underlying nuclear interactions.
	
	Despite these underlying microscopic differences, the radially reweighted sampling ($\text{ARM}_r$) reproduces the target pair-distance distributions of both \textit{ab-initio} models with high fidelity across the full range $0\leq\Delta r\leq5\,\mathrm{fm}$. Figure~\ref{fig_A1}(b) shows the corresponding normalized pair distributions $g'(\Delta r)$. After normalization, the two distributions exhibit a common overall shape, indicating that the bulk pair-distance distribution is predominantly determined by the common single-particle 3pF radial profile, while the model-dependent nuclear correlations primarily modify the pair distribution at small separations.

	\bibliography{ref}
	
\end{document}